# Altitude Dependence of Atmospheric Temperature Trends: Climate Models vs Observation

David H. Douglass[1]*, Benjamin D. Pearson[1] and S. Fred Singer[2]

1. Dept of Physics and Astronomy, University of Rochester, Rochester, NY 14627
2. Science & Environmental Policy Project and University of Virginia, Charlottesville, VA 22903

## Abstract

As a consequence of greenhouse forcing, all state-of-the-art general circulation models predict a positive temperature trend that is greater for the troposphere than the surface. This predicted positive trend increases in value with altitude until it reaches a maximum ratio with respect to the surface of as much as 1.5 to 2.0 at about 200-400 hPa. However, the temperature trends from several independent observational data sets show decreasing as well as mostly negative values. This disparity indicates that the three models examined here fail to account for the effects of greenhouse forcings.

*corresponding author. douglass@pas.rochester.edu

## I. Introduction

All state-of-the-art general circulation models (GCMs) predict a warming trend due to various greenhouse forcings that is greater for the troposphere than the surface. The calculated warming trend reaches a maximum at an altitude corresponding to a pressure of 200-400 hPa and decreases to negative values in the stratosphere. These model results disagree with many observations of the lower troposphere over the last 25 years, which show the surface to be warming faster than the air just above.

Jones et al. (1999) reported that the surface warms by 190 [$10^{-3}$K/decade] relative to the satellite data [referring to a mean altitude of ~2 km] (Christy et al. 2000). Gaffen *et al.* (2000), in a study of the tropics that included radiosonde measurements, reached similar conclusions. Singer (2001) discussed resolving the disparity as the most important research priority.
Christy (2001) has done a comparative study of trend-lines between Hadley sea-surface temperatures and four temperature sets of the air above the sea surface. He finds that "[T]he difference in temperature trends is statistically significant in the tropical belt...." Christy et al. (2001) have extended the study of the tropics to include buoys which had thermometers 3m in the air and 1m below the surface. They find statistically significant cooling of the air relative to the water. Chase et al. (2004) compared four models with five data sets and considered the possibility that variability in the models could account for the disparity. They conclude that "model variability and uncertainty in applied forcings cannot produce the recently observed tropospheric temperature characteristics."

This disparity between state-of the-art models and observations is even more dramatic when the comparison is extended to the mid- and upper troposphere, and is the subject of this paper. We show below that the various models all predict that greenhouse effects are stronger at mid- to high- tropospheric altitudes. The maximum temperature trend, which can be twice that at the surface, occurs between 400 and 200 hPa. Above this pressure/altitude the effect decreases and trends become negative when one reaches the stratosphere.

We consider the most recent results of three state-of-the-art global climate models: Hadley CM3, DOE PCM, and GISS SI2000 for the satellite era, the last 20 to 25 years. These models consider a variety of different climate forcings: greenhouse gases, sulfate aerosols, ozone -- and natural forcings, such as volcano and solar. We show that the models yield a positive trend line, ~200-300 [$10^{-3}$K/decade], which increases with altitude with any combination of forcings. For this same time interval we consider four observational data sets: surface temperature (ST), microwave sounding units satillite (MSU), balloon-borne radiosonde, and NCAR/NCEP Reanalysis (NNR).

## II. Model Results and Data

### *A. Model Results*

### 1. Hadley CM3

Tett et al. (2002) have given the latest results for the Hadley CM3 computer model. This is a coupled atmospheric/ocean general circulation model. They consider four ensembles of external forcings;1. Natural. The simulations were forced with the solar irradiance and the stratospheric aerosol loading from explosive volcano eruptions..2. GHG. Historic changes in well-mixed greenhouse gases.3. Tropo-Anthro. GHG plus anthropogenic sulfur emissions and their implied changes to cloud albedos, plus tropospheric ozone.4. Anthro. As in tropo-anthro, except that depletion of stratospheric ozone was included.

Numerical results for these four simulations vs. latitude at pressures [altitude] 1000, 850, 700, 500, 300, 200, 100 hPa were made available to us by Tett (2003).The zonal average was computed at each latitude. Results for the 'Anthro' simulation are shown in Fig 1a. The plots for GHG and Tropo-Anthro were similar. The same results plotted vs pressure/altitude are shown in Figures 2a, 2b, 2c, 2d for GHG and Anthro.All of the modeled trends are positive except for the 200 hPa level, where there is a negative section beyond about 60N. This is due to the fact that the 200 hPa level includes not only the troposphere, but also extends into the stratosphere. The trends have a much higher magnitude in the tropics.

## 2. DOE PCM model

This Parallel Coupled Model (PCM) [Meehl et al. 2003] is a fully coupled global ocean-atmosphere-sea ice-land surface model without flux adjustments. Five different external climate forcings were considered: greenhouse gases, sulfate aerosol (direct effect only), stratospheric and tropospheric ozone, solar, volcano.  Various combinations of forcings were employed, with simulations vs. latitude at pressures [altitude] 1000, 850, 700, 500, 400, 300, 250, 200, 100 hPa. Of particular interest is the "ALL" experiment [Santer et al. 2003] in which all five forcings were present.  The numerical results for the ALL experiment for the period 1979-1999 were made available to us by Meehl and Arblaster (2003).The zonal average trend vs. latitude is shown in Fig 1B. The same data vs pressure/altitude are shown in Figures 2a. 2b, 2c. 2d.

## 3. GISS SI2000 Simulation

The Goddard Institute for Space Studies (GISS) GISS SI2000 atmospheric model is a representative general state-of-the-art climate model (Hansen et al 2002), using 12 vertical layers.  They consider different model oceans; ocean A uses SSTs and sea ice; ocean B includes a deep ocean.  They have considered a sequence of six forcings: 1) well-mixed greenhouse gases; 2) stratospheric aerosols; 3) solar irradiance; 4) ozone; 5) stratospheric water vapor; and 6) tropospheric aerosols. We use the model results for the case of the six forcings for ocean A (shown in their fig. 16) for the interval 1979-98.  We plot the values for Global, Northern Hemisphere, Tropics, Southern Hemisphere vs pressure/altitude in Figures 2a, 2b, 2c,2d.

### *B. Observational Data: Four data sets*

We obtain observational data from MSU, ST, and NNR and Radiosondes.  We have chosen the time interval 1979 to 1996 for our analysis, but have also ascertained that our conclusions are not substantially affected by extending the interval to 2002 (Douglass et al. 2004).

### 1. Surface data ST

The surface temperature anomalies are taken from Jones et al. [1999] in a 5º by 5º grid.  We note for later reference that sea-surface temperatures (SST) are based mainly on measurements from underneath the surface.  There are missing data near the poles and at high southern latitudes, and also at other places.

### 2. Satellite-MSU

The lower troposphere temperature anomalies are taken from measurements from microwave sounding units (MSU; channel 2, t2lt [~2 km]) on polar-orbiting satellites. The surface is uniformly sampled, with some 40,000 measurements/day.  (Christy et al. 2000)

### 3. Radiosondes HadRT2.0

Radiosondes carried on weather balloons measure climate parameters vs altitude, such as temperature. The HadRT2.0 radiosonde data set is described by Parker et al. (1997).  It gives temperature anomalies at standard pressure levels (850-50 hPA).  Brown et al. (2000) have computed trend lines from this data set for the period 1979-1998.  We plot these trend-line values listed in their Table 1 for global, northern and southern hemispheres and the tropics at pressure levels 850, 700, 500, 300, 200, 150, 100 hPa in Figures2a, 2b, 2c, 2d.

## 4. NCAR/NCEP Reanalysis (NNR)

We have also used data from a NCEP/NCAR Reanalysis (NNR) that produces a retroactive record of more than 50 years of global analysis of atmospheric fields (Kisteler et al. 2001). The reanalysis program takes inputs from many measurement systems, such as radiosondes, COADS, and computes the values of various climate variables at standard times and spatial intervals using the reanalysis model. It yields monthly values of temperature for each grid cell at various standard altitudes. We note that NNR data do not use surface data from Hadley/Jones. NNR also does not use Christy *et al.* MSU data, but do utilize some raw MSU data which are calibrated weekly by radiosondes. Thus the temporal temperature evolution of NNR is independent of Christy et al.'s MSU data. We plot temperatures trend-lines at p= 850, 700, 500, 200, 100, and one designated tmp.2m, the temperature at the altitude of "2 meters", to which we assign the value p=1000 hPa.In our altitude plots (Fig 2) we have indicated the position and range [standard deviation from the average] in pressure for the tropopause, as follows.
NH: 150-200hPa;  Tropics: 100-115;  SH: 140-170;  G: 120-160.
The latitude dependence of the pressure at the tropopause was taken from <http://cdc.noaa.gov/>

### *C. Latitude Dependence*

The latitude dependence of the HadCM3 and DOE PCM models is shown in Figures 1a, 1b. At the surface the trend-line is positive for all latitudes, with the largest values occurring in the tropics. As one goes to higher altitudes (lower pressures), the trend-lines increase until one reaches ~300hPa, beyond which they decrease. Fig 1C shows the MSU, NCAR/NECP, and ST observational data. The MSU, NNR [and Radiosonde not shown] are seen to generally disagree with the models except in the high-latitude Northern Hemisphere. The ST data are not in disagreement with the models. This disparity between ST and other observations is discussed in Douglass et al. (2004)

### *D. Altitude Dependence*

The differences between the three models and the four observational data sets become apparent when one computes zonal averages. Fig 2 shows averages for Global, Tropics, Northern and Southern hemispheres vs. pressure/altitude.

**Global Average:** Fig2a is the global plot. All three models for any of the scenarios [not all of the scenarios are shown] show a value ~180 to 220 [$10^{-3}$ K/decade] at p~1000 hPa, which increases with altitude, maximizing at about 400-300 hPA and decreasing at higher altitudes. However, the MSU, NCEP, and radiosonde observations show temperature trend values around zero at p = 1000 hPa and do not increase with altitude. The ST data point is consistent with the models.

**Tropical (30S – 30N):** **Fig 2b** is similar to Global. But the calculated temperature trends maximize at 250 hPa and are positive, while observed trends are negative and decrease even further beyond about 300 hPa. Again the models agree with ST.

**Southern Hemisphere (30 – 60S):** Fig 2c is very similar to Global and Tropics. Maximum of models occurs at 400-300hPa. Here MSU, NNR, and radiosonde data have values ~ -50 to -150 [$10^{-3}$K/decade] at p ~ 1000hPa and increase in value to ~ 0 to 50 at about p ~500 to 300. Again the models agree with ST.

**Northern Hemisphere (30 – 60N): Fig 2d.** The models show a trend value of ~200 to 300 [$10^{-3}$K/decade] out to 400 to 500 hPa and then decrease. The observational near-surface values of ST, MSU and NNR.2m appear to agree with the models. The radiosondes and the NNR values at higher altitudes lie below the models.

### III. Analysis

**Models**

All three of the selected state-of-the-art climate model calculations show the same altitude dependence in all of the zonal plots for similar forcing scenarios. They give positive trend-lines at the earth's surface in all latitude bands. The values generally increase with altitude showing maximum trends at pressures of the order of 300 hPa.

**Observational data**

For the near-surface data (ST, MSU, and NNR.2m) there is a disparity of ST with other data sets: MSU, NNR.2m. The disparity for globally averaged temperature trends has been documented by NRC (2000). Douglass *et al.* (2004) showed that this near-surface disparity was not in the NH band but occurred mainly in the tropics. Here [in this paper] we find that the near-surface trends (ST, MSU, NNR.2m) disagree in the global, tropical, and SH bands but agree in the NH band --- consistent with Douglass *et al*. (2004)

The mid- to high- troposphere data sets (radiosondes and NNR -- other than NNR.2m) generally agree in all latitude bands. In general, the trend-lines start with low values and decrease to negative values with increasing altitude. [Christy (2004) has noted that because of a change in radiosonde instrumentation any trend-lines derived from radiosondes and NNR (other than NNR.2m) show low values over Germany and Russia; this may affect the plots for the NH (30N-60N).]

**Comparison of Models with the Observational Data**

In the comparison of the models with the near surface data (ST, MSU, and NNR.2m) we find that the models all generally disagree in the Global, tropics, and SH averages with MSU and NNR.2m. The models agree with ST, MSU, NNR in the NH, and with ST in all zonal averages. For the mid- to high- troposphere data (NNR and radiosondes) the models disagree in all zones.

### IV. Discussion and Conclusions:

We have found that while the models generally agree with each other, they disagree with the observations. In particular, the three state-of-the-art greenhouse models (Hadley, DOE PCM, and GISS SI2000) examined here show positive temperature trends that increase with altitude, reaching values greater than the near-surface trends by as much as 50 to 100 percent. However, the existing observational data sets show decreasing as well as mostly negative trends since 1979. We are therefore faced with two alternatives.

1. **The models are correct and account for all relevant forcings.** If so, then we must conclude that the observational data sets -- MSU, NNR and Radiosondes -- are all incorrect.

2. **The models do not fully capture the multitudinous climate effects (including various feedbacks) of an increase in greenhouse gases.** Since the observed surface temperature trends (ST) agree with the models, then they too must be questioned.

It seems improbable that results from satellites (MSU), NCAR/NCEP reanalysis (NNR), and Radiosondes, which agree with each other, would all be wrong. Therefore, it seems more likely that both the models and observed surface trends are problematic. Their apparent agreement may be a coincidence or perhaps reflect a "tuning" of the models to the surface temperature trends.

**Acknowledgements**
This research was supported in part by the Rochester Area Research Foundation. We wish to thank Simon Tett for the Hadley model CM3 numerical results, Gerald Meehl and Julie Arblaster for the DOE PCM numerical results, and R. S. Knox for valuable discussions. Additional thanks to V. Patel and Yi-Lun Ding for doing some of the computations.

**Figure Captions**

Figure 1. Zonal averages trend-line [$10^{-3}$K/decade] vs. latitude for various pressure levels (altitude).
   a. HadCM3 model. ANTHRO [ghg+sulfates+tropo ozone+strato ozone]. [(1985-1995) minus (1961-1980)]
   b. DOE PCM model. "ALL" = historic ghg+sulfate+ozone+solar+volcano simulations. 1979-1999]
   c. Observational data [1979-1996]: MSU-troposphere [equiv. pressure = 770 hPa]; ST-surface (Hadley); and NNR: 2m, 700 hPa, 200 hPa.

Figure 2. Temperature trend-line [$10^{-3}$K/decade] versus log pressure (altitude) for different zonal averages. Observations [filled symbols, solid lines]: MSU, ST (Hadley), NNR, HadRT2.0 radiosondes. Models [open symbols, dotted lines]: Hadley model CM3, DOE PCM, and GISS SI2000 . Tropopause range is shaded. Tropopause data can be found at <http://cdc.noaa.gov/>
   a. Global average. All of the data sets have an average taken from 90S to 90N with exception of ST (Hadley), whose average was taken from 67.5S to 67.5N.
   b. Tropics. (30S to 30N)
   c. Southern Hemisphere (60S to 30S).
   d. Northern Hemisphere. (30N to 60N)

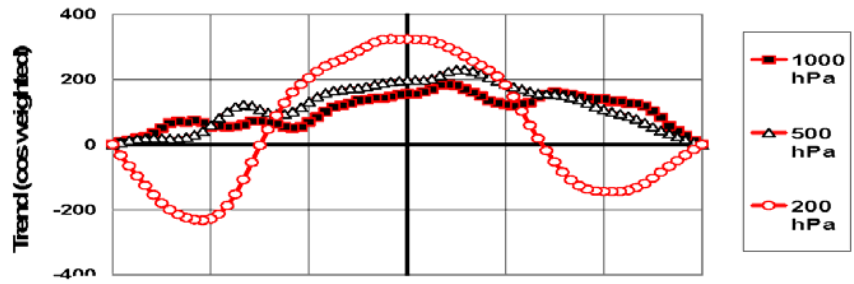
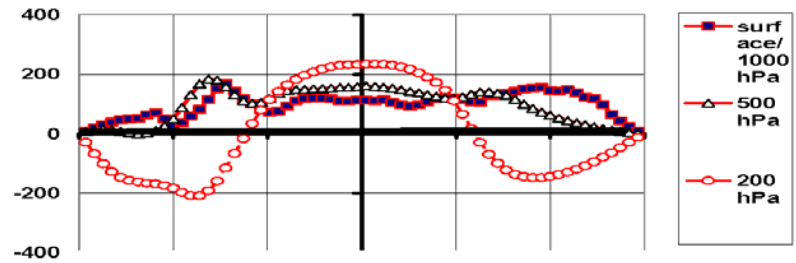
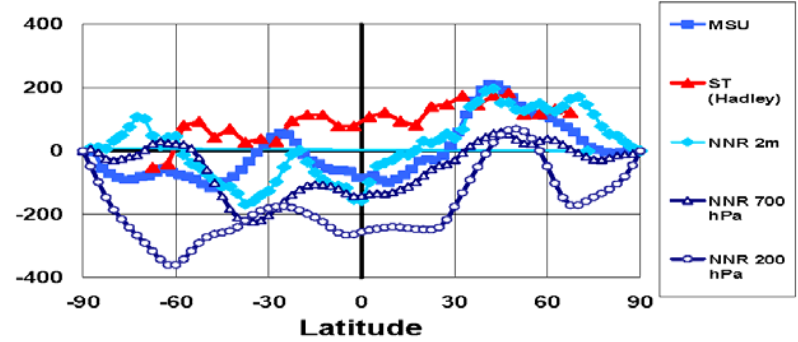

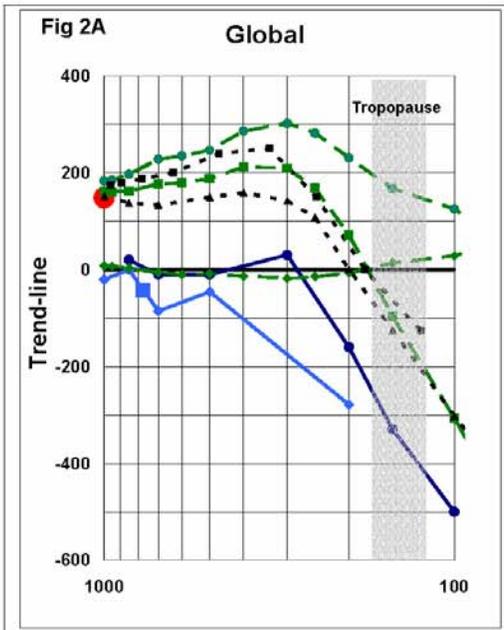
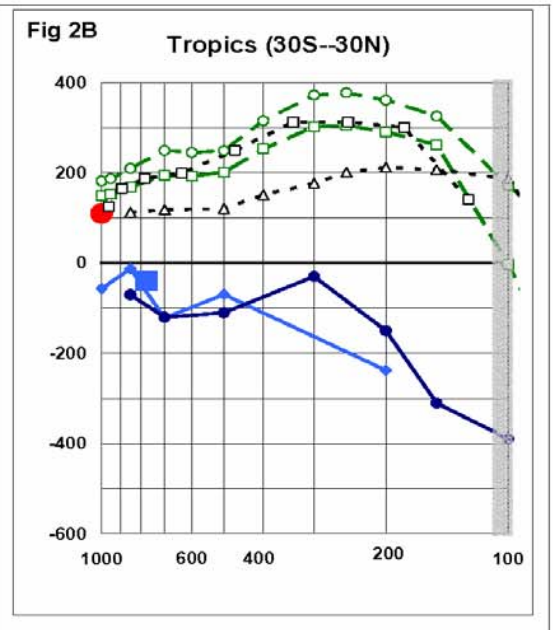
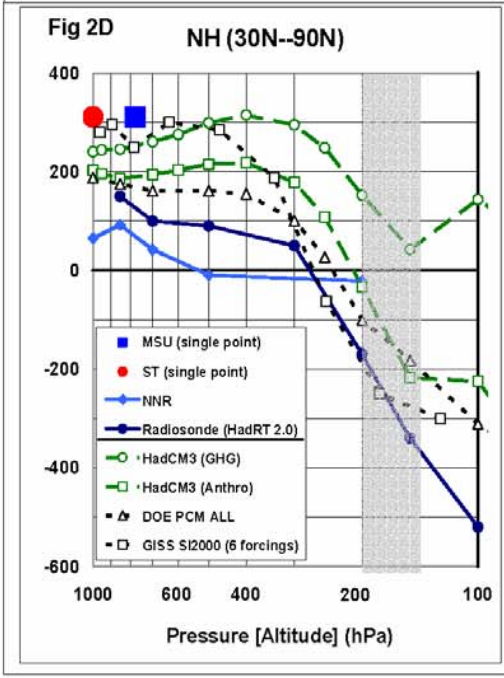
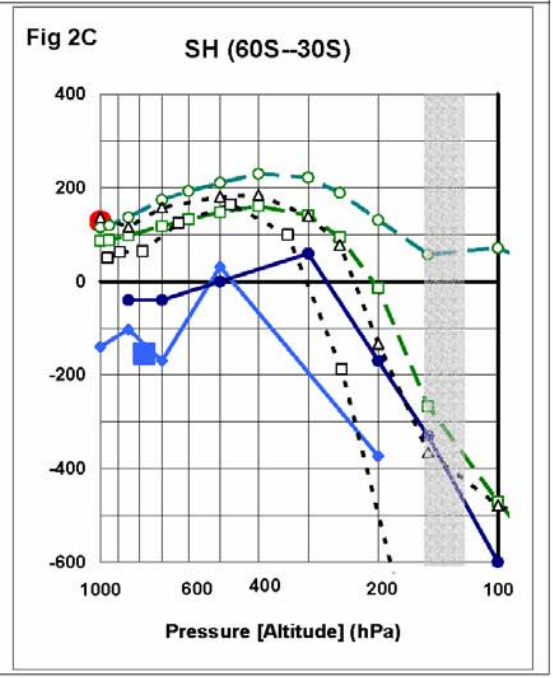